# A System-level Engineering Approach for Preliminary Performance Analysis and Design of Global Navigation Satellite System Constellations

Marco Nugnes[1], Camilla Colombo[1], Massimo Tipaldi[2]

**Abstract** – *This paper presents a system-level engineering approach for the preliminary coverage performance analysis and the design of a generic Global Navigation Satellite System (GNSS) constellation. This analysis accounts for both the coverage requirements and the robustness to transient or catastrophic failures of the constellation. The European GNSS, Galileo, is used as reference case to prove the effectiveness of the proposed tool. This software suite, named GNSS Coverage Analysis Tool (G-CAT), requires as input the state vector of each satellite of the constellation and provides the performance of the GNSS constellation in terms of coverage. The tool offers an orbit propagator, an attitude propagator, an algorithm to identify the visibility region on the Earth's surface from each satellite, and a counter function to compute how many satellites are in view from given locations on the Earth's surface. Thanks to its low computational burden, the tool can be adopted to compute the optimal number of satellites per each orbital plane by verifying if the coverage and accuracy requirements are fulfilled under the assumption of uniform in-plane angular spacing between coplanar satellites.*

## Nomenclature

| Symbol | Description |
|---|---|
| $\alpha$ | Right ascension [deg] |
| $\alpha_G$ | Greenwich meridian right ascension [deg] |
| $\alpha_T$ | Temperature gradient [K/m] |
| $\beta$ | Ballistic coefficient [m$^2$/kg] |
| $\gamma$ | Pitch angle [deg] |
| $\delta$ | Declination [deg] |
| $\Delta r_\odot$ | Satellite relative distance w.r.t. the Sun [km] |
| $\Delta t$ | Time interval [s] |
| $\varepsilon_0$ | Void dielectric constant [F/m] |
| $\phi$ | Satellite longitude [deg] |
| $\lambda$ | Wavelength [m] |
| $\kappa$ | Geometric signal bending [m] |
| $\mu_\oplus$ | Earth gravitational parameter [km$^3$/s$^2$] |
| $\eta$ | Antenna efficiency |
| $\vartheta$ | Azimuthal angle [deg] |
| $\rho$ | Atmospheric density [kg/m$^3$] |
| $\psi$ | Roll angle [deg] |
| $\nu$ | Angular spacing [deg] |
| $\boldsymbol{\omega}$ | Satellite angular velocity [rad/s] |
| $\omega_p$ | Pericentre anomaly [deg] |
| $\omega_\oplus$ | Earth angular velocity [rad/s] |
| $\Omega$ | Right ascension of the ascending node [deg] |
| $\zeta$ | Yaw angle [deg] |
| $a$ | semi-major axis [km] |
| $\mathbf{A}$ | Rotation matrix |
| $\mathbf{b}$ | Leverage vector w.r.t. the pole [m] |
| $\mathbf{B}$ | Magnetic field vector [T] |
| BER | Bit Error Rate |
| $c$ | Speed of light [m/s] |
| $c_a$ | Absorption coefficient |
| $c_d$ | Diffuse reflection coefficient |
| $c_s$ | Specular reflection coefficient |
| $C_D$ | Drag coefficient |
| $d$ | Antenna diameter [m] |
| $d_{dry}^z$ | Dry tropospheric delay [m] |
| $d_{tropo}$ | Tropospheric delay [m] |
| $d_{wet}^z$ | Wet tropospheric delay [m] |
| $\mathbf{D}$ | Atmospheric drag per unit mass [N/kg] |
| $e$ | Water vapour pressure [mbar] |
| $e_s$ | Surface water vapour pressure [mbar] |
| $E_b$ | Energy per bit |
| EIRP | Equivalent Isotropic Radiated Power [dB] |
| $f$ | Signal frequency [Hz] |
| $F$ | Solar radiation force [N] |
| $F_e$ | Solar radiation power per unit surface [W/m$^2$] |
| FCC | Federal Communications Commission |
| FSL | Free Space Losses [dB] |
| G-CAT | GNSS Coverage Analysis Tool |
| $G_{R_x}$ | Receiver antenna gain [dB] |
| $G_{T_x}$ | Transmitter antenna gain [dB] |
| GEO | Geostationary Orbit |
| GNSS | Global Navigation Satellite System |
| GPS | Global Positioning System |
| $H$ | Signal bandwidth [Hz] |
| $i$ | Orbital inclination [deg] |
| $i_{iono}$ | Ionospheric delay [m] |
| ITU | International Telecommunication Union |
| $J_2$ | Second zonal coefficient |
| $L$ | Distance between the receiver and satellite [m] |
| $L_{gas}$ | Gas attenuation [dB] |
| $L_{rain}$ | Rain attenuation [dB] |
| $L_{T_x}$ | Transmitter line losses [dB] |
| LEO | Low Earth Orbit |
| $\mathbf{J}$ | Inertia matrix |
| $k'_2$ | Refractivity coefficient [K/mbar] |
| $k_3$ | Wet refractivity coefficient [K$^2$/mbar] |
| $k_B$ | Boltzmann constant [J/K] |
| $k_d$ | Derivative gain [Hz] |
| $k_p$ | Proportional gain |
| $m$ | Satellite mass [kg] |
| $M$ | Mean anomaly [deg] |
| $n_s$ | Satellite mean angular velocity [rad/s] |
| $N_0$ | Signal noise power [dB] |
| $p$ | Semi-latus rectum [km] |
| $\mathbf{p}_m$ | Residual magnetic induction |
| $P$ | Atmospheric pressure [mbar] |
| $P_s$ | Surface pressure [mbar] |
| $P_{T_x}$ | Transmitter power [dB] |
| $q$ | Elemental charge [C] |
| $\mathbf{r}$ | Relative position vector [km] |
| $\mathbf{R}$ | Inertial satellite position vector [km] |
| $R_\oplus$ | Earth equatorial radius [km] |
| $R_d$ | Dry air specific gas constant [J/KgK] |
| $S$ | Satellite surface [m$^2$] |
| $\mathbf{S}_0$ | Static moment [kgm] |
| $\mathbf{T}$ | Disturbing torque [Nm] |
| $T_e$ | Equivalent noise temperature [K] |
| $T_m$ | Mean temperature of water vapour [K] |
| $T_s$ | Surface temperature [K] |
| TEC | Total Electron Content |
| $\mathbf{u}$ | Control action [Nm] |
| $\mathbf{v}_{in}$ | Satellite inertial velocity vector [km/s] |
| $\mathbf{v}_{rel}$ | Satellite relative velocity vector [km/s] |
| $V$ | Magnetic field potential |
| WGS | World Geodetic System |
| $z'$ | Zenith angle at the ionospheric point [deg] |

## I. Introduction

Satellite constellations signed the beginning of a new era in terms of space mission design philosophy [1]. While in the second half of the last century monolithic spacecraft were deployed to accomplish the mission requirements, from the beginning of the new millennium the tendency changed to allocate the instrumentations and the tasks of space (scientific) missions over more than one spacecraft. The list of the future space missions involving satellite constellations demonstrates why it is useful to develop engineering approaches for the analysis of their performance.

One of the most demanding missions exploiting telecommunication satellite systems has been recently proposed by SpaceX to shorten the communication time among internet users on Earth and space-faring satellites, speeding up surfing speeds. In November 2018, SpaceX received the approval for launching an additional bunch of 7518 satellites reaching the number of 11943 satellites [2] and the first 240 satellites were already injected in orbit. SpaceX is not the only company aiming to provide internet broadband services to individual users. In March 2018, OneWeb asked the approval to authorise 1260 more satellites to be added to the approved 720 ones [3], while Telesat received approval for a 117 satellite network in September 2018 [4]. Another mission involving 78-108 Low Earth Orbit (LEO) satellites is LeoSat, which is addressed to enterprises, employing big data transactions with lower latency. Indeed, this is a service that Geostationary Orbit (GEO) satellites cannot provide and the company plans to enter this new market [5]. The Federal Communications Commission (FCC) which oversees the regulations regarding telecommunication systems is adapting their rules to the incoming constellation deployment. Indeed, it gave the approval to SpaceX under the condition of full constellation deployment in nine years. If SpaceX fails to reach full deployment in that time, its authorised number of satellites will shrink to the number already in orbit [6]. This means that in the next decades several satellite constellations will be deployed.

Besides communication and internet services, satellite constellations have been exploited also for Earth observation through remote sensing techniques (for instance, the study of the ionosphere total electron content parameter [7] and the global ocean altimetry [8]).

This paper addresses Global Navigation Satellite System (GNSS) constellations, whose space segment consists of a group of navigation satellites in orbit around the Earth placed in an orbit positioning to get a minimum requirement for global coverage. The ground segment involves the ground stations and antennas and it is responsible for managing the constellation of navigation satellites, controlling the core functions of the navigation missions as well as determining the integrity information [9]. The user segment consists of a receiver converting the navigation signal of the satellites to position and velocity at a specific time. Another application involving GNSS is security. Indeed, the international entities for the transportation sector exploit the high precision determination offered by GNSS constellations to air traffic management to prevent the occurrence of accidents.

In literature, the GNSS performance analysis has been addressed by examining specific engineering aspects. For instance, Sünderhauf et al. [10] developed a method to mitigate the multipath effect with the aim of improving the GNSS positioning determination of a generic user. Multipath effects consist in the repetition of the same signal shifted in time due to the reflection of the signal itself coming from different paths [11]. This phenomenon is more dangerous in large cities, where the reflection of the navigation signals is considerably higher. Coulot et al. [12] proposed an optimisation method of the GNSS performance by analysing the on-ground reference stations. Their work relies on the employment of algorithms to develop the future ground stations network to improve the communication with the GNSS space segment and, consequently, the performance of the system. Pan et al. [13] investigated the GNSS positioning performance in partially obstructed environments. By using a carrier double-difference model, they analysed the influence of the satellite-pair geometry on the correlation among the carrier phase observation equations, and proposed a method for determining the position of the reference satellites by using the minimum condition number rather than the maximum elevation. Tadic et al. [14] presented a web-based GNSS performance analysis and simulation tool offering a cross comparison module to benchmark devices from an established reference. Their work focuses on the modelling of the signal propagation and uses statistical data collected from existing GNSS. All the mentioned works addressed the GNSS constellation performance by considering the enhancement of only one single aspect of these complex systems, such as the navigation signal multipath phenomenon.

This paper presents a system-level engineering methodology and the corresponding tool named GNSS Coverage Analysis Tool (G-CAT) to perform the design of GNSS or generic satellite constellations, by means of a coverage performance analysis starting from the initial orbital parameters of the space segment. The tool can suit different types of pointing scenarios (like geocentric, geodetic or a generic pointing altered by the effect of orbital and attitude perturbations), and gives the possibility to analyse not only the nominal configuration, but also scenarios where one or more satellites fail both for a transient period or permanently [15]. Finally, the robustness to failures associated with the accuracy in the position determination is a driver for the optimisation of the number of satellites per each orbital plane by running the tool few numbers of times.

This manuscript is an extended version of [16], and explores more deeply all the aspects of the proposed methodology, together with the models embedded and their assumptions. The paper is organised as follows. In Sec. II the G-CAT functional model is described with its inputs and outputs. Sec. III presents the modelling of each function contained in the G-CAT functional model.

Sec. IV describes the G-CAT graphical user interface. Sec. V presents the results obtained in case of Galileo and Global Positioning System (GPS) nominal operational scenarios and in some failure conditions. Sec. VI shows how to optimise the number of satellites per each orbital plane. Finally, Sec. VII concludes the paper and outlines the future directions.

## II. G-CAT Functional Model Overview

This section introduces the G-CAT functional model shown in the block diagram of Fig. 1. G-CAT contains a section dedicated to the space segment, a section related to the signal propagation, and one for the coverage analysis.

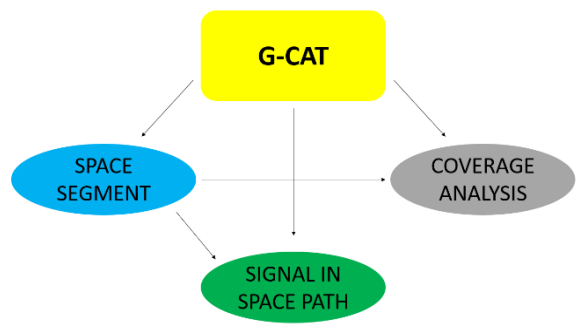

Fig. 1. G-CAT functional model block diagram

### II.1. Space Segment

The space segment block consists of an attitude and orbit propagator. The orbit propagator works by taking as input the initial state vector of each satellite of the constellation in terms of classic orbital elements and a propagation time expressed in hours. A standard two-body problem propagator or a perturbed propagator can be chosen.

An attitude propagator is also available, which takes as input all the physical properties and the initial conditions for the Euler angles and the angular velocities of each spacecraft of the constellation. The attitude propagator will be presented in Sec. III as it is important to understand its modelling and the type of output provided to the other functions. Attitude perturbations are already embedded in the models, and the propagation time is equal to the time used for the coverage analysis.

### II.2. Signal in Space Path

The second functional block deals with the signal propagation starting from the navigation antenna up to the ground surface. All the models embedded in the tool rely on approximations of the main factors affecting the navigation signal propagation. The signal propagation requires the modelling of the interaction of an electromagnetic wave with the atmosphere, precisely two layers: the ionosphere and the troposphere.

Several phenomena like refraction, scintillation, and multipath need to be considered. For the sake of simplicity, only the refraction is considered by the tool. Both the ionosphere and the troposphere can cause the refraction of the signal for different reasons, with the result of the signal deviation from the straight path and the introduction of a delay in the reception of the signal. The ionosphere is responsible for the deflection of the signal due to the presence of charged particles which interact with the electromagnetic wave. On the other side, the troposphere also introduces a refraction and a curvature in the signal propagation, which is related to the interaction with atmosphere molecules like water vapour.

### *II.3. Coverage Analysis*

The last functional block determines the coverage region of each satellite given the spacecraft position vector and the direction of the navigation line of sight. This information is derived, respectively, from the orbit propagator, whose output is the spacecraft position vector in Cartesian coordinates, and from the attitude propagator. These two inputs are provided by two different propagators since it is assumed that the orbit propagation is decoupled from the satellite attitude. To determine the coverage region of each satellite, the coverage analysis functional block uses a new refined analytical approach, which models the Earth's surface as an oblate ellipsoid of rotation [17].

Such block also contains the function in charge of the verification of the constellation coverage. Starting from the knowledge of each coverage area associated to a single spacecraft, a function verifies whether the points located on a grid computed from the Earth's surface are within or not the coverage area and counts the number of satellites visible from the regions on the Earth's surface. This way, it is possible to derive the coverage and the degree of accuracy according to the number of visible satellites. The coverage analysis gives as output four coverage indices:

- "Red index", i.e., the number of points on the grid covered by less than 4 satellites.
- "Yellow index", i.e., the number of points on the grid covered by exactly 4 satellites.
- "Green index", i.e., the number of points on the grid covered by more than 4 satellites.
- "Global index", i.e., the number of points on the grid covered by at least 4 satellites and it is equal to the sum of the "yellow index" and "green index".

These indices have been introduced by considering the resolution of the general global positioning problem, also known as trilateration process. Starting from the measurement of the pseudo-distance between a satellite and a receiver, a single measurement provides the position of the user on the surface of a sphere. Two measurements reduce the size of the problem into a planar one since the user's position this time is on the region described by a circle which is the intersection of the two spheres having radii equal to the pseudo-distances between the satellite and the receivers. Finally, three measurements give as output the intersection of two circles, which allows choosing the right position between two points. Even if this ambiguity can be solved, the number of measurements is not enough due to the presence of the time-bias caused by the different accuracy between the on-board satellite's clock and the receiver's clock. The result is that four measurements are needed to derive unequivocally the user's position on the Earth's surface [18].

## III. G-CAT Functional Model Description

In this section, the modelling approach of the G-CAT functional models is presented. Each function can be configured via specific panels of the G-CAT graphical user interface, which will be presented in Sec. IV.

### *III.1. Orbit Panel Functions*

The "orbit panel" belongs to the functional block of the space segment. The default orbit propagator implements the standard two-body Keplerian orbit propagator, but the user has the possibility to select an orbit propagator which considers perturbations.

The default propagator models in a simplified way the two most important perturbations for GNSS constellations: the non-sphericity of the Earth and the solar radiation pressure. Atmospheric drag can be neglected since the altitude of a generic GNSS is above 28000 km. For future constellations (e.g., Starlink, OneWeb), for which the altitude is under 1000 km, the effect of the atmosphere will be included.

The non-sphericity of the Earth is modelled by considering only the secular effect of the $J_2$ zonal harmonic on the right ascension of the ascending node, the pericentre anomaly, and the mean anomaly. The formulations for the secular rates of the orbital parameters are [19]:

$$\dot{\Omega}_{\text{sec}} = -\frac{3n_s R_\oplus^2 J_2}{2p^2} \cos(i) \quad (1)$$

$$\dot{\omega}_{p_{\text{sec}}} = \frac{3n_s R_\oplus^2 J_2}{4p^2}\left(4 - 5\sin^2(i)\right) \quad (2)$$

$$\dot{M}_{\text{sec}} = -\frac{3n_s R_\oplus^2 J_2}{4p^2}\left(3\sin^2(i) - 2\right) \quad (3)$$

where $\dot{\Omega}_{\text{sec}}$, $\dot{\omega}_{p_{\text{sec}}}$, $\dot{M}_{\text{sec}}$ are the secular rates of the right ascension of the ascending node, the pericentre anomaly, and the mean anomaly respectively, $n_s$ represents the mean motion of the satellite, $R_\oplus$ is the Earth mean equatorial radius, $J_2$ is the second zonal harmonic coefficient, $p$ is the semi-latus rectum and $i$ is the orbital inclination. The secular rates are used to determine the linear variation with time of the corresponding orbital parameters:

$$\Omega = \Omega_0 + \dot{\Omega}_{sec} \Delta t \tag{4}$$

$$\omega_p = \omega_{p_0} + \dot{\omega}_{p_{sec}} \Delta t \tag{5}$$

$$M = M_0 + \dot{M}_{sec} \Delta t \tag{6}$$

The solar radiation pressure is modelled by considering only the effect on the semi-major axis [19]:

$$\dot{a} = -\frac{2a^2}{\mu_\oplus} F \Delta r_\odot \tag{7}$$

with $a$ and $\dot{a}$ representing the semi-major axis and its rate respectively, $\mu_\oplus$ is the Earth gravitational constant, $\Delta r_\odot$ is the satellite relative distance with respect to the Sun, and $F$ is the solar radiation force. Also in this case, the semi-major axis rate is used to determine the linear variation with time of the semi-major axis. This kind of variation is not secular, but it is a cyclic perturbation with zero net effect in one solar year if the satellite is assumed to be always in sunlight:

$$a = a_0 + \dot{a}\Delta t \tag{8}$$

The default perturbed orbital propagator is quite simple. Its execution does not require a significant computational load. All the satellites are assumed as concentrated in their centre of mass, therefore the orbit propagation can be decoupled from the attitude. The time scales for the orbit propagation of the satellite are extremely larger than the ones used for the attitude propagator. In Sec. IV, the advantage of the decoupling between the attitude and the orbit propagator is explained for both the short and long period analysis of the constellation.

### III.2. Attitude Propagator

The attitude propagator is one of the most important functions of the tool since it allows to generalise the direction of the navigation signal and to simulate more realistic orbital scenarios. For this reason, a brief description of the modelling functions and related assumptions will be provided. In this case, each satellite is modelled as a rectangular parallelepiped rigid body. The attitude representation adopted for the constellation is the one using Euler angles and direct cosine matrices. Euler angles are used for the inputs to the propagator while all the processing is performed with the direct cosine matrices since they present the advantages to be more robust to numerical integration and do not present relevant singularities. The block diagram showed in Fig. 2 summarises the data flow within the attitude propagator.

The starting point of the attitude propagator is the resolution of the dynamics through the Euler equations which needs as input the disturbances, given by the attitude perturbations, and the control actions of the actuators together with the initial angular velocity.

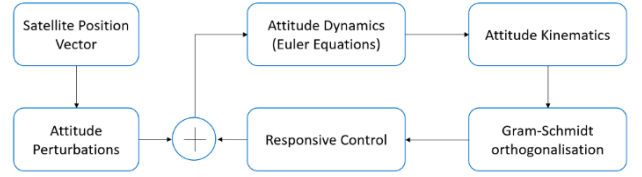

Fig. 2. Attitude propagator functional block diagram

The result of the dynamics is the angular velocity at the current time that is combined with the initial attitude configuration to solve the attitude kinematics in terms of direct cosine matrices. The attitude state in terms of direct cosine matrix and angular velocity is known at the current time and the corrective control action to track the desired state is implemented. This control action is added to the satellite attitude perturbations which depend only on the Cartesian position of the spacecraft and the next cycle begins solving again the Euler equations with the updated variables.

From the knowledge of the initial conditions in terms of angular velocity $\boldsymbol{\omega} = \left[\omega_x, \omega_y, \omega_z\right]$, the Euler equations [20] are solved to compute the value of the angular velocity at each instant with respect to the Earth-centred inertial frame:

$$\begin{cases} J_{xx}\dot{\omega}_x = (J_{yy} - J_{zz})\omega_y\omega_z + u_x \\ J_{yy}\dot{\omega}_y = (J_{zz} - J_{xx})\omega_x\omega_z + u_y \\ J_{zz}\dot{\omega}_z = (J_{xx} - J_{yy})\omega_x\omega_y + u_z \end{cases} \tag{9}$$

with $J_{xx}$, $J_{yy}$, $J_{zz}$ representing the diagonal terms of the inertia matrix J expressed in the principal axes reference frame of the satellite and $u_x$, $u_y$, $u_z$ are the components of the torques generated from the attitude perturbations and spacecraft control actuators.

After having computed the attitude dynamics, the attitude kinematics can be retrieved. This is done by solving the following differential equation [21]:

$$\dot{\mathbf{A}}_{B/L} = -[\boldsymbol{\omega}_{B/L}]^\wedge \mathbf{A}_{B/L} \tag{10}$$

where $\mathbf{A}_{B/L}$ is the direct cosine matrix expressing the rotation matrix from the satellite orbital frame coincident with the local frame and denoted with the letter "*L*", and the satellite body frame denoted with the letter "*B*", while the $[\boldsymbol{\omega}_{B/L}]^\wedge$ is the skew-symmetric matrix built from the knowledge of the angular velocity of the spacecraft orbital frame expressed in the body frame:

$$[\boldsymbol{\omega}_{B/L}]^\wedge = \begin{pmatrix} 0 & -\omega_z & \omega_y \\ \omega_z & 0 & \omega_x \\ -\omega_y & \omega_x & 0 \end{pmatrix} \quad (11)$$

The direct cosine matrix $\mathbf{A}_{B/L}$ is a classic "321" rotation matrix [22]:

$$\mathbf{A}_{B/L} = \begin{pmatrix} \cos(\gamma)\cos(\zeta) & \cos(\gamma)\sin(\zeta) & -\sin(\gamma) \\ -\sin(\zeta)\cos(\psi) + \cos(\zeta)\sin(\gamma)\sin(\psi) & \cos(\zeta)\cos(\psi) + \sin(\zeta)\sin(\gamma)\sin(\psi) & \cos(\gamma)\sin(\psi) \\ \sin(\zeta)\sin(\psi) + \cos(\zeta)\sin(\gamma)\cos(\psi) & -\cos(\zeta)\sin(\psi) + \sin(\zeta)\sin(\gamma)\cos(\psi) & \cos(\gamma)\cos(\psi) \end{pmatrix}$$

where $\psi$, $\gamma$ and $\zeta$ are the roll, pitch and yaw rotation angles of the body reference system with respect to the local reference frame.

The angular velocity vector $\boldsymbol{\omega}_{B/L}$ is obtained as the result of two contributions:

$$\boldsymbol{\omega}_{B/L} = \boldsymbol{\omega}_{B/N} - \mathbf{A}_{B/L}\boldsymbol{\omega}_{L/N} \quad (12)$$

where $\boldsymbol{\omega}_{B/N}$ and $\boldsymbol{\omega}_{L/N}$ are the angular velocity vectors of the body reference frame and local reference frame with respect to the inertial reference system denoted with the letter "*N*".

One of the disadvantages in using direct cosine matrices for solving the attitude kinematics is the loss of the orthogonality property due to numerical errors after the numerical integration. This problem can be solved by using a Gram-Schmidt algorithm [23] which makes the matrix orthogonal at each instant. Another problem associated with "321" rotation matrices is the singularity occurring for values of the pitch angle equal to $90°$ or $270°$. However, such values for the pitch angle are never achieved under the action of the attitude control system and so they can be neglected in the propagation processing.

The last model worth mentioning is the one related to the control action, $\mathbf{u}$. The formulation used is the following [24]:

$$\mathbf{u} = -k_d \mathbf{J}\boldsymbol{\omega}_e - k_p \mathbf{J}\left(\mathbf{A}_e^T - \mathbf{A}_e\right)^\vee - \mathbf{J}[\boldsymbol{\omega}_e]^\wedge \mathbf{A}_e \boldsymbol{\omega}_{L/N} \\ + \boldsymbol{\omega}_{B/N} \wedge \mathbf{J}\boldsymbol{\omega}_{B/N} \quad (13)$$

with $k_d$ and $k_p$ representing the derivative and proportional gain respectively while $\boldsymbol{\omega}_e$ and $\mathbf{A}_e$ are the angular velocity and attitude matrix error with respect to the desired state. The default desired state is the configuration where the spacecraft is aligned along the navigation antenna line of sight, coincident with the *z*-axis of the body reference frame. For this reason, the desired angular velocity should allow only rotations around the *z*-axis and the desired attitude matrix is the one associated with a yaw angle equal to zero.

This type of control is called responsive control since it has the property to cancel the first term in the Euler equations (9) and to make the dynamics resolution faster. Of course, this method presents also the disadvantage to make slower the process if the values of the inertia matrix are not accurate in their determination [24]. After this processing, the *z*-direction of the body reference frame, which is assumed to be aligned with the navigation signal direction, is determined and in this way, it is possible to simulate the moving line of sight. The attitude propagator models four kinds of perturbations: atmospheric drag, gravity gradient, solar radiation pressure and magnetic torque.

(14)

The atmospheric drag is the reactive force related to the atmosphere nearby the spacecraft. Even if the density is very low with respect to the Earth's surface level, there is a small contribution. This effect becomes insignificant for altitudes higher than 10000 km. To determine the value of the torque, the expression of the drag resistance per unit mass is written:

$$\mathbf{D} = \frac{1}{2}\beta\rho v_{rel}^2 \hat{\mathbf{v}}_{rel} \quad \text{with} \quad \beta = \frac{SC_D}{m} \quad (15)$$

where *m* is the mass of the spacecraft, $C_D$ is the drag coefficient which is in around 2-3 for common spacecraft, *S* is the wet area of the spacecraft, $\mathbf{v}_{rel}$ is the velocity vector relative to the atmosphere, $\rho$ is the atmospheric density and $\beta$ is the ballistic coefficient, parameter that characterises the response of the spacecraft to drag. In G-CAT it is required to insert the value of the ballistic coefficient which is known from the design and geometry of the satellite, while the density is computed from the height of the satellite through an exponential model, and the relative velocity is derived from the knowledge of the inertial velocity as follows:

$$\mathbf{v}_{rel} = \mathbf{v}_{in} - \boldsymbol{\omega}_\oplus \wedge \mathbf{R} \quad (16)$$

The inertial velocity, $\mathbf{v}_{in}$, and the position vector in Cartesian coordinates, $\mathbf{R}$, come from the orbital propagator while the $\boldsymbol{\omega}_\oplus = [0, 0, 2\pi/86400]$ is the angular velocity of the Earth in the geocentric equatorial frame neglecting the nutation and precession of the rotational axis. The torque associated to the atmospheric drag is, finally, given by:

$$\mathbf{T}_{drag} = \mathbf{b} \wedge \mathbf{D} \qquad (17)$$

where **b** is the vector associated to the leverage between the application point of the force and the pole. The leverage is assumed to be 10% of the length of the spacecraft.

The gravity gradient is the difference in terms of gravity acceleration of different parts of the spacecraft since the spacecraft is not considered as a dot point but as a 3D rigid body. To derive the expression of the gravity gradient it is necessary to express first the torque acting on an infinitesimal element of mass, $dm$.

$$d\mathbf{T} = -\mathbf{r} \wedge \frac{\mu_\oplus dm}{\|\mathbf{r}+\mathbf{R}\|^3}(\mathbf{r}+\mathbf{R}) \qquad (18)$$

where **r** is the relative position vector of the infinitesimal mass with respect to the body reference frame, while **R** is the position vector of the spacecraft centre of mass. The total torque acting on the spacecraft is the integral over the whole body of the infinitesimal torque:

$$\mathbf{T} = -\int_B \mathbf{r} \wedge \frac{\mu_\oplus dm}{\|\mathbf{r}+\mathbf{R}\|^3}(\mathbf{r}+\mathbf{R}) \qquad (19)$$

It is possible to linearise the expression of the denominator since the magnitude of **r** is much smaller than **R**. Taking the series expansion of the denominator in the integral close to zero:

$$\|\mathbf{r}+\mathbf{R}\|^{-3} \approx \mathrm{R}^{-3}\left(1 - 3\frac{\mathbf{R}\cdot\mathbf{r}}{\mathrm{R}^2}\right) \qquad (20)$$

Substituting the linearised form in Eq. (19):

$$\mathbf{T} = -\frac{\mu_\oplus}{\mathrm{R}^3}\int_B \mathbf{r} \wedge \left(1 - 3\frac{\mathbf{R}\cdot\mathbf{r}}{\mathrm{R}^2}\right)(\mathbf{r}+\mathbf{R})dm \qquad (21)$$

Considering the first addendum:

$$\int_B \mathbf{r} \wedge (\mathbf{r}+\mathbf{R})dm = -\mathbf{R} \wedge \mathbf{S}_0 \qquad (22)$$

where $\mathbf{S}_0$ is the static moment of the spacecraft. The first cross product has result equal to 0 and in the second term the position vector **R** is independent from the infinitesimal mass and it can be taken outside the integration. If the reference system of the body frame has the origin in the satellite's centre of mass like the one considered, the static moment is equal to 0. The other term in Eq. (21) gives the final expression of the gravity gradient torque:

$$\mathbf{T} = \frac{3\mu_\oplus}{\mathrm{R}^5}\int_B (\mathbf{r}\cdot\mathbf{R})(\mathbf{r}\wedge\mathbf{R})dm \qquad (23)$$

This integral can be further developed. Indeed, considering the body reference frame, those vectors assume these expressions:

$$\mathbf{r} = x\hat{\mathbf{r}} + y\hat{\boldsymbol{\theta}} + z\hat{\mathbf{k}} \qquad (24)$$
$$\mathbf{R} = c_1\hat{\mathbf{r}} + c_2\hat{\boldsymbol{\theta}} + c_3\hat{\mathbf{k}} \qquad (25)$$

where $[c_1, c_2, c_3]$ are the direction cosines of the radial direction. Substituting inside Eq. (23):

$$\mathbf{T} = \frac{3\mu_\oplus}{\mathrm{R}^3}\int_B (xc_1 + yc_2 + zc_3)\begin{pmatrix} yc_3 - zc_2 \\ zc_1 - xc_3 \\ xc_2 - yc_1 \end{pmatrix} dm \qquad (26)$$

Evaluating all the terms under the integral sign, and considering that the reference frame consists of principal axes of inertia:

$$\mathbf{T} = \frac{3\mu_\oplus}{\mathrm{R}^3}\begin{pmatrix} (J_{zz} - J_{yy})c_2c_3 \\ (J_{xx} - J_{zz})c_1c_3 \\ (J_{yy} - J_{xx})c_1c_2 \end{pmatrix} \qquad (27)$$

Therefore, if one of the principal axes is aligned with the radial direction the torque is zero because only one of the direction cosines is non-zero.

Solar radiation illuminating the outer surface of a satellite generates a pressure, that in turn generates a force and a torque around the centre of mass of the satellite. The main sources of electromagnetic radiation are the direct solar radiation, the solar radiation reflected by the Earth (or by any other planet) and the radiation directly emitted by the Earth. Solar radiation intensity varies with the inverse square distance from the source, so that direct solar radiation is almost constant for geocentric orbits, independent from the orbit radius since the distance to the Sun does not change significantly. On the contrary, the reflected radiation and the Earth radiation intensity are strongly dependent on the orbit radius. The average pressure due to radiation can be evaluated as:

$$P = \frac{F_e}{c} \qquad (28)$$

where $c$ is the speed of light and $F_e$ is the solar radiation power per unit surface. Radiation forces can be modelled assuming that part of the incident radiation is absorbed, part is reflected in a specular way and part reflected with diffusion. Coefficients $c_a$, $c_d$, $c_s$ are respectively the coefficient of absorption, diffuse reflection and specular reflection, and are constrained by the following equality:

$$c_a + c_d + c_s = 1 \quad (29)$$

Merging the three components due to the three possible effects, the total force due to radiation is evaluated as:

$$\mathbf{F}_j = -PS_j\left(\hat{\mathbf{O}}\cdot\hat{\mathbf{N}}_j\right)\left[(1-c_s)\hat{\mathbf{O}} + 2\left(c_s\hat{\mathbf{O}}\cdot\hat{\mathbf{N}}_j + \frac{2}{3}c_d\right)\hat{\mathbf{N}}_j\right] \quad (30)$$

with $\hat{\mathbf{O}}$ denoting the unit direction of the satellite-Sun vector and $\hat{\mathbf{N}}$ the unit direction of the normal to each spacecraft surface $j$. The total torque will be the summation of the single torque on each spacecraft surface.

$$\mathbf{T} = \sum_{j=1}^{6} \hat{\mathbf{r}}_j \wedge \mathbf{F}_j \quad (31)$$

Again, the leverage of the solar radiation pressure has been considered 10% of the length of the spacecraft. The last source of possible attitude perturbations is offered by the interference of the geomagnetic field with the electronics onboard of the spacecraft. The expression for the magnetic torque is:

$$\mathbf{T} = \mathbf{p}_m \wedge \mathbf{B} \quad (32)$$

where $\mathbf{p}_m$ is the residual magnetic induction due to parasitic currents in the satellite and $\mathbf{B}$ is the Earth's magnetic field. Normally, $\mathbf{p}_m$ is an undesired effect while $\mathbf{B}$ is always somehow present. There are models that, given the satellite position, allow evaluating its components. Magnetic torques on a satellite therefore do not depend on in its inertia properties but rather on its position and attitude. So, the next open point is how the magnetic field is evaluated. The magnetic field $\mathbf{B}$ can be modelled in a precise way as the gradient of a scalar potential $V$, that is $\mathbf{B} = -\nabla V$. Normally, $V$ is modelled as series expansion of spherical harmonics [21]:

$$V(r,\vartheta,\phi) = R_\oplus \sum_{n=1}^{k}\left(\frac{R_\oplus}{r}\right)^{n+1}\sum_{m=0}^{n}\left(g_n^m \cos(m\phi) + h_n^m \sin(m\phi)\right)P_n^m(\vartheta) \quad (33)$$

where $B_r$ represents the radial component, positive outwards, $B_\vartheta$ the co-elevation component, positive if directed towards south, and $B_\phi$ the azimuth component, positive towards east. Coefficients $g_n^m$ and $h_n^m$ are evaluated under the assumption that polynomials $P_n^m$ are normalised according to Schmidt [21]:

$$\int_0^\pi \left[P_n^m(\vartheta)\right]^2 \sin(\vartheta)d\vartheta = \frac{2(2-\delta_m^0)}{2n+1} \quad (34)$$

where $\delta_m^0$ is the Kronecker delta, $\delta_i^j = 1$ if $i=j$ or $\delta_i^j = 0$ if $i \neq j$. In alternative, a normalisation due to Gauss [21] is possible, according to which:

$$P_n^m = S_{n,m} P^{n,m} \quad (35)$$

$$S_{n,m} = \left[\frac{(2-\delta_m^0)(n-m)!}{(n+m)!}\right]^{\frac{1}{2}} \frac{(2n-1)!}{(n-m)!} \quad (36)$$

Coefficients $S_{n,m}$ are independent from $r, \vartheta$ and $\phi$ can be evaluated from the Gaussian coefficients $g_n^m$ and $h_n^m$, following the rule:

$$g^{n,m} = S_{n,m} g_n^m$$
$$h^{n,m} = S_{n,m} h_n^m$$
$$S_{0,0} = 1$$
$$S_{n,0} = S_{n-1,0}\frac{(2n-1)}{n} \quad \text{for} \quad n \geq 1$$
$$S_{n,m} = S_{n,m-1}\left[\frac{(\delta_m^1+1)(n-m+1)}{(n+m)}\right]^{\frac{1}{2}} \quad \text{for} \quad m \geq 1$$

In a similar way, it is possible to get recursive formulas for the polynomials $P^{n,m}$:

$$P^{0,0} = 1$$
$$P^{n,n} = \sin(\vartheta)P^{n-1,n-1}$$
$$P^{n,m} = \cos(\vartheta)P^{n-1,m} - K^{n,m}P^{n-2,m}$$
$$K^{n,m} = 0 \quad \text{for} \quad n = 1$$
$$K^{n,m} = \frac{(n-1)^2 - m^2}{(2n-1)(2n-3)} \quad \text{for} \quad n > 1$$

With these models, the components of the magnetic field, $\mathbf{B}$, in the $[r,\vartheta,\phi]$ reference frame are:

$$B_r = \sum_{n=1}^{k}\left(\frac{R_\oplus}{r}\right)^{n+2}(n+1)\sum_{m=0}^{n}\left(g^{n,m}\cos(m\phi) + h^{n,m}\sin(m\phi)\right)P^{n,m}(\vartheta) \quad (37)$$

$$B_\vartheta = -\sum_{n=1}^{k}\left(\frac{R_\oplus}{r}\right)^{n+2}\sum_{m=0}^{n}\left(g^{n,m}\cos(m\phi)\right.$$
$$\left.+h^{n,m}\sin(m\phi)\right)\frac{\partial P^{n,m}(\vartheta)}{\partial \vartheta} \quad (38)$$

$$B_\phi = -\frac{1}{\sin(\vartheta)}\sum_{n=1}^{k}\left(\frac{R_\oplus}{r}\right)^{n+2}\sum_{m=0}^{n}m\left(-g^{n,m}\sin(m\phi)\right.$$
$$\left.+h^{n,m}\cos(m\phi)\right)P^{n,m}(\vartheta) \quad (39)$$

In a geocentric inertial reference frame, the components of the magnetic field assume the form:

$$B_x = \left(B_r\cos(\delta) + B_\vartheta\sin(\delta)\right)\cos(\alpha) - B_\phi\sin(\alpha) \quad (40)$$

$$B_y = \left(B_r\cos(\delta) + B_\vartheta\sin(\delta)\right)\sin(\alpha) + B_\phi\cos(\alpha) \quad (41)$$

$$B_z = B_r\sin(\delta) - B_\vartheta\cos(\delta) \quad (42)$$

where $\delta$ is the satellite declination $(\delta = \pi/2 - \vartheta)$ and $\alpha$ is the right ascension, linked to the longitude of the satellite by the relation:

$$\alpha = \phi + \alpha_G \quad (43)$$

with $\alpha_G$ the Greenwich meridian right ascension.

### III.3. Failure Function

This function allows turning "on" or "off" the navigation signal of a satellite for a given period. Indeed, it is possible to shut down the satellite for the whole simulation or to disable it for a transient period, and then turn it on again.

This gives the possibility to simulate realistic scenarios, such as the execution of recovery procedures for the navigation payload and its navigation signal. During this time interval, the satellite is not operating properly, and it should be considered as not existing in the constellation. Another possible failure scenario is the shutdown of the navigation payload due to under-voltage or over-voltage problems. In such cases, the navigation payload operational status switches from the nominal mode into a safe mode, where it stays until the problem has been solved. The failure function gives the possibility to consider many cases where the outage of the navigation signal occurs and to check how the performance of the GNSS constellation is affected.

### III.4. Coverage Function

The coverage function is in charge of determining the coverage area where a satellite is visible (see Fig. 3), by starting from the satellite position vector and the navigation signal direction and modelling the Earth's surface as a sphere or an oblate ellipsoid of rotation according to the World Geodetic System (WGS-84) representation [25].

The reader can refer to [17] for more details on the equations used to compute the coverage area for an oblate ellipsoid of rotation. Such equations simplify to the ones of a perfectly spherical Earth if the values of the equatorial and polar radius of the Earth are the same.

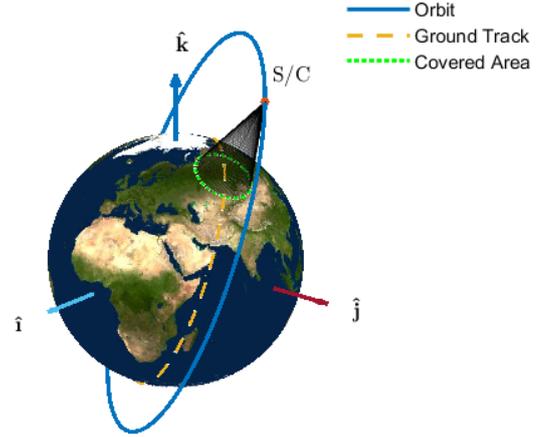

Fig. 3. Geometry for the field of view in the Earth-centred reference frame

### III.5. Signal Propagation Function

The tool offers the possibility to have a first preliminary estimation of the link budget starting from the signal properties and the antenna's characteristics. Moreover, it gives as output a figure of merit on the delay in the signal propagation due to the interaction with the atmosphere. It is important to highlight that these computations are decoupled from the coverage analysis since they are external and add simply more information.

The link budget is one of the most important quantities during the design phase of a space mission giving information about the quality of the telecommunication signal. For navigation satellites where the signal represents the payload of the mission, it becomes of crucial importance. The link budget results as output of a series of operations the signal to noise ratio, starting from a fixed set of variables like carrier frequency, antenna gain, spacecraft height, and internal hardware. Once the signal to noise ratio is evaluated, according to the type of codification and modulation used for the signal, the quality of the communication is verified. The starting point of the link budget computation is the carrier frequency, $f_{car}$, of the navigation signal which must be approved from the International Telecommunication Union (ITU). Then, according to the type of antenna onboard and the hardware it is possible to evaluate the Equivalent Isotropic Radiated Power (EIRP) summarising the quality of the signal exiting from the transmitter. The EIRP is evaluated using the following formula [26]:

$$EIRP = P_{T_x} + G_{T_x} - L_{T_x} \qquad (44)$$

where $P_{T_x}$ is the power of the transmitter, $G_{T_x}$ is the antenna gain and $L_{T_x}$ represents the line losses associated to the transmitter. It has to be highlighted that in the previous formula all the quantities are expressed in decibel ($dB$). The value of the antenna gain $G_{T_x}$ is fixed once the physical properties of the antenna have been established:

$$G_{T_x} = \left(\frac{\pi d}{\lambda}\right)^2 \eta \qquad (45)$$

with $d$ representing the diameter of the antenna, $\lambda$ the wavelength of the signal and $\eta$ the antenna efficiency. This quantity is non-dimensional and must be converted in $dB$ in order to be used:

$$G_{T_{x_{dB}}} = 10 \log_{10} G_{T_x} \qquad (46)$$

Once the EIRP is evaluated, the next step is to compute the losses existing for the distance between the user receiver on the Earth's surface and the antenna of the satellite. This kind of losses are called Free Space Losses (FSL) and may be estimated using the following formulation [26]:

$$FSL = \left(\frac{\lambda}{4\pi L}\right)^2 \qquad (47)$$

where $L$ represents the distance between the user receiver and the antenna. Also, in this case the conversion in $dB$ is necessary. At this stage the signal arrives at the boundary of the atmosphere and so new losses must be introduced. For this purpose, some statistical plots modelling the value of the loss associated to an atmospheric phenomenon exist. The atmospheric losses may be distinguished in two main components: gas losses and rain losses. The gaseous losses are associated to the resonance of the frequency of the signal with the frequency of vibration of the molecules in the atmosphere, in particular molecular oxygen, $O_2$, and water vapour, $H_2O$. It is important to select a value for the carrier far away from the absorption peaks as it is shown in Fig. 4.

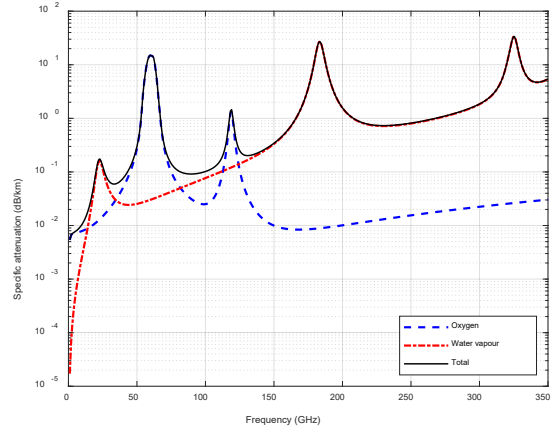

Fig. 4. Specific attenuation losses due to atmospheric gases versus the navigation signal frequency

The second source is rain and for each location on the Earth a certain probability of rain in one year is evaluated. This way, it is possible to have a figure of merit for the expected loss in one year as shown in Fig. 5. Both the models for the specific attenuation due to gases and rain have been taken from [27].

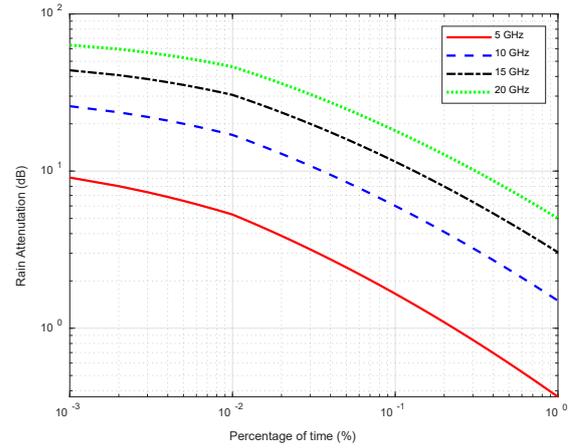

Fig. 5. Specific attenuation losses due to rain versus the percentage of rain time for different navigation signal frequencies

The last part in the signal to noise ratio computation is to evaluate the properties of the receiving antenna. Also, in this case there is a gain that can be computed in the same way done for the transmitter. However, an additional source of disturbance is present which is represented by the noise power which depends on the bandwidth and on the physical temperature of the system. Generally, an equivalent noise temperature $T_e$ for each component of the receiver is introduced to express noise power. The temperatures are summed to get an equivalent system noise temperature in such way that the noise power is expressed as:

$$N_0 = k_B H T_e \qquad (48)$$

with $k_B$ the Boltzmann constant equal to $1.3806 \times 10^{-23}$ J/K and $H$ the bandwidth of the signal. The final quantity that is evaluated is the energy per bit to noise power ratio, $E_b/N_0$, which is strictly related to the signal to noise ratio [26]:

$$\frac{E_b}{N_0} = EIRP - FSL - L_{rain} - L_{gas} + G_{R_x} - N_0 \quad (49)$$

At this stage according to the type of modulation, the Bit Error Rate (BER) is derived as function of the energy per bit to noise power ratio following [28]. The BER is an index of the quality of the signal as it is shown in Fig. 6. Indeed, in order to have a good communication it is expected that the BER is at least $10^{-5}$ bit/s.

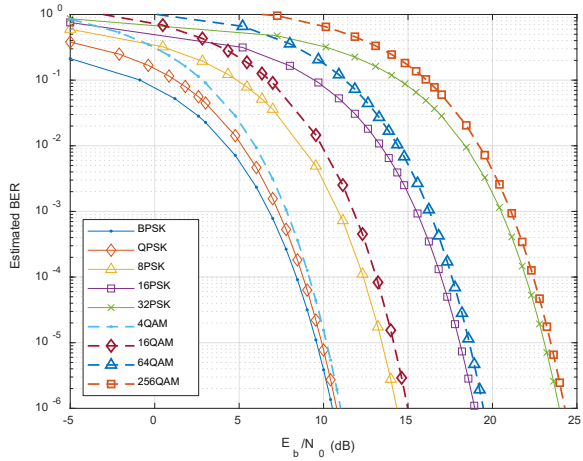

Fig. 6. Estimated Bit Error Rate (BER) versus signal over noise ratio for different typed of codification

The modelling of the ionosphere is performed using first-order approximations taken from [29]. The ionosphere is the upper part of the atmosphere where charged particles are mixed with neutral particles. The charged particles are created by photoionisation caused by incoming UV and X-radiation from the Sun: gas molecules are heated, and electrons are liberated then.

The rate of this ionisation depends on the density of gas molecules and the intensity of the radiation. In the neutral atmosphere charged particles are practically absent, since the created charged particles are recombined rapidly due to the high density of particles. In the ionosphere, however, only the charged particles can influence the propagation of radio waves. Mainly, the free electrons affect the propagation since the free ions are much heavier than the electrons.

The interaction of the ionosphere with the navigation signal can be decomposed into four sources of delay called first-order, second-order, third-order delay, and geometrical signal bending denoted respectively as $i^{(1)}$, $i^{(2)}$, $i^{(3)}$ and $\kappa$ [29]. The second and third-order delays are often referred to as the ionospheric higher-order terms. The total ionospheric delay can be written as:

$$i_{iono} = i^{(1)} + i^{(2)} + i^{(3)} + \kappa \quad (50)$$

All the formulations for the evaluation of the different sources for the ionospheric delay are taken from [29]. The first-order delay, $i^{(1)}$, which is also the most significant for the ionospheric delay computation is approximated as follows:

$$i^{(1)} = \frac{A}{2f^2} TEC \quad \text{with} \quad A = \frac{q^2}{4\pi^2 m_e \varepsilon_0} \quad (51)$$

where $f$ is the frequency of the navigation signal, $TEC$ is the Total Electron Content representing the integral of the electron density in the ionosphere, $q = 1.602 \times 10^{-19}$ C is the elemental charge, $m_e = 9.109 \times 10^{-31}$ kg is the electron mass and $\varepsilon_0 = 8.854 \times 10^{-12}$ F/m is the void dielectric constant. The second-order delay, $i^{(2)}$, is associated to geomagnetic field:

$$i^{(2)} \approx \frac{qA}{f^3 2\pi m_e} \|\mathbf{B}\| \cos(\vartheta) TEC \quad (52)$$

with $\|\mathbf{B}\|\cos(\vartheta)$ representing the projection of the magnetic field vector along the signal path. The third-order delay, $i^{(3)}$, can be formulated as follows:

$$i^{(3)} \approx \frac{3A^2}{8f^4} \tau N_{e,\max} TEC \quad (53)$$

where $N_{e,\max}$ is the maximum electron density in the ionosphere and $\tau$ represents a shape factor that is equal to 0.66. Finally, the geometric signal bending is evaluated by using the following approximation:

$$\kappa \approx \frac{A^2}{8f^4} \tan^2(z') \eta N_{e,\max} TEC \quad (54)$$

with $z'$ representing the zenith angle at the ionospheric point, that is the intersection between the line of sight connecting the satellite with the user receiver and the ionosphere boundary modelled as a thin-uniform shell located at 350 km from the Earth's surface.

The tropospheric delay represents the time shift introduced by the neutral part of the atmosphere due to the introduction of a geometric curvature of the navigation signal because of the variation in the refraction index in the different layers of the troposphere. The modelling of the tropospheric delay is carried out also in this case by decomposing it into two main terms in order to distinguish the effects due to the dry gases (primarily nitrogen and oxygen) and the water vapour. The formulation of the dry tropospheric delay is given by [30]:

$$d_{hyd}^z = \frac{10^{-6} k_1 R_d}{g_m} P_s \qquad (55)$$

where $k_1 = 77.604$ K/mbar is the refractivity constant of the dry gases, $R_d = 287.054$ J/(KgK) is the specific gas constant for dry air, $g_m$ is the gravity at the centre of mass of the column of air and $P_s = 1013.25$ mbar is the surface pressure. The second part of the tropospheric delay which is denoted as *wet delay* can be evaluated with the following equation [30]:

$$d_{wet}^z = \frac{10^{-6}\left(k_2' + k_3/T_m\right) R_d}{g_m (\Lambda + 1)} e \qquad (56)$$

$$T_m = T_s\left(1 - \frac{\alpha_T R_d}{(\Lambda+1)g_m}\right) \qquad e = e_s\left(\frac{P}{P_s}\right)^{(\lambda+1)}$$

with $T_m$ the mean temperature of water vapor, $k_2' = k_2 - k_1(M_w/M_d) = 16.52$ K/mbar a modified refractivity coefficient, $k_3 = 3.776 \times 10^5$ K$^2$/mbar the wet refractivity coefficient, $\Lambda$ an empiric coefficient depending on the local conditions, $\alpha_T$ the temperature gradient with respect to the surface height, $e_s = 11.7$ mbar the surface water vapour pressure, $T_s$ the surface temperature, $e$ the water vapour pressure and $P$ the tropospheric pressure that can be computed using the exponential model of the standard atmosphere. The values of the coefficients used in G-CAT are $\alpha_T = 6 \times 10^{-3}$ K/m and $\Lambda = 3$. The total tropospheric delay is get from the summation of the two contributions:

$$d_{tropo}^z = d_{hyd}^z + d_{wet}^z \qquad (57)$$

The expressions mentioned before are valid for the zenith directions. If the direction is different from the zenith, the effect in the new direction is obtained by considering mapping functions that require further modelling and they are not implemented in G-CAT.

## IV. G-CAT Graphical User Interface and Configuration

In the following section, the G-CAT user graphical interface (see Fig. 7) is presented. The procedures to configure the G-CAT tool, run a simulation, and analyse the related results are also provided.

The G-CAT tool configuration consists in defining all the parameters needed to set up a specific simulation scenario. The first step is to insert the initial state vector in terms of orbital elements for each satellite of the constellation. The tool can handle up to 30 satellites at the current time (in line with the number of satellites of existing GNSS), and a tab is dedicated to each satellite.

After the selection of the type of orbit propagator, the propagation time can be inserted. This time can be different from the coverage analysis time. This way, the orbit and attitude propagations are performed as part of the coverage performance analysis. More specifically, the two propagations are decoupled and carried out for different integration times. Indeed, the orbit perturbations time scales are much larger than the attitude ones. This means that the user can start an orbit propagation for a longer time and the tool gives as output the evolution of the orbit parameters with each time instant. At this stage, the user can change the initial orbital parameters selected with the one obtained from the orbit propagation. This way, it is possible to simulate the attitude behaviour for a small time interval by starting from the orbital parameters of the constellation at a different epoch to check the effects of the orbital perturbations on the performance.

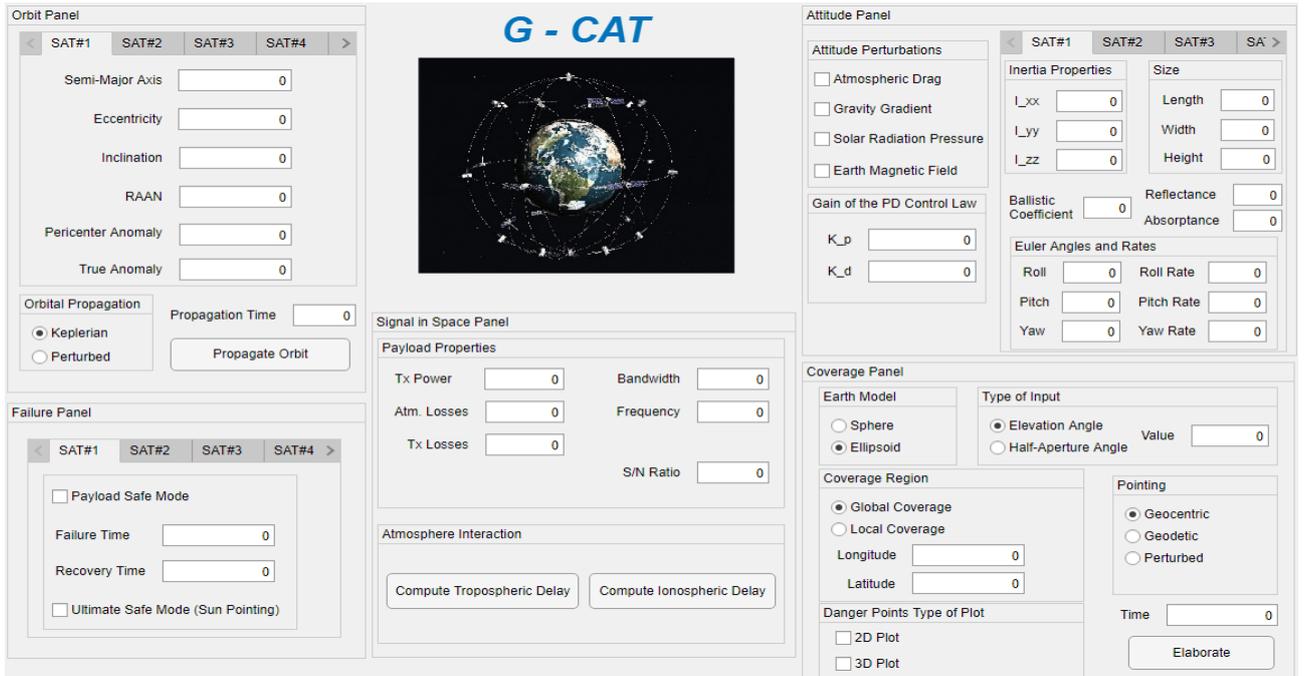

Fig. 7. G-CAT graphical user interface

Once the orbit propagation is concluded, the attitude panel can be filled in the same way of the orbital one. For each satellite a tab is available where the geometrical characteristics can be inserted, together with the initial attitude conditions:
- Principal moment of inertia
- Length, width and height, assuming that each satellite is a parallelepiped
- Ballistic coefficient
- Reflectance and absorptance coefficients
- Gains for the control law
- Initial Euler angles and angular velocities.

There is also the possibility to select the type of attitude perturbations to be included in the propagation.

The failure panel is a box where the user can indicate whether a satellite is in nominal or in safe mode, and in the latter case, the duration of such off-nominal condition as explained in Sec. III.

The signal in space panel is another secondary box divided in "payload properties", where the information about the navigation signal should be filled to compute the figures of merit for the signal to noise ratio, and the "atmosphere interaction", where the tropospheric and ionospheric delays are computed. As already mentioned in Sec. III, the evaluations of the two delays are decoupled from the coverage analysis because the G-CAT tool calculates a first-order estimation for those variables.

Finally, in the coverage panel, the user can choose the type of modelling for the Earth's surface (sphere or oblate ellipsoid of rotation), the input for the navigation signal, the type of pointing, the set of points to be analysed, and the type of output. The navigation signal is modelled as a conical field of view by having as axis of symmetry the navigation signal direction. For this reason, the coverage area is derived by considering either the conical half-aperture angle or the minimum elevation angle from where the satellites are visible on the Earth's surface [17]. A small panel offers the possibility to perform both the local and the global coverage analysis. In the first case, the user can insert the geographic coordinates of the location whose coverage has to be verified. In the latter case, the tool discretises the Earth's surface as a grid, and verifies if every point on such a grid is covered. The user has also the possibility to choose three different types of pointing, according to the mission requirements: geocentric, geodetic and perturbed. The geocentric pointing has the satellite always looking towards the centre of the Earth. The geodetic pointing has the satellite pointing along the local normal direction, which is different from the geocentric direction in the oblate ellipsoid case. Finally, the perturbed pointing starts from an initial direction, and then considers the evolution over time of the navigation signal direction caused by the attitude perturbations. Only in the case of perturbed pointing selection the attitude propagator is involved in the coverage analysis together with the orbit propagator. Indeed, the other two pointing modes imply a fixed direction for the navigation antenna and no attitude propagation is required.

The user can also select if the output of the analysis should be plotted on a 3D representation of the Earth's globe or on a planar one. The G-CAT tool flags with a red cross all the regions on the Earth's surface which are not covered for at least one time instant during the whole simulation. The last parameter to be set before the tool can start the simulation is the time for the coverage performance. Simulations are not time-consuming processes and can last from less than 5 minutes (in case of geocentric or geodetic pointing) up to 10 minutes (in case of perturbed pointing) using an Intel® Core™ i7-4710HQ CPU @2.50 GHz.

Fig. 8 shows how the G-CAT functions are executed when the coverage performance analysis is launched. The tool receives as input all the initial data regarding the orbit and attitude configuration of each satellite of the constellation. The user can decide to analyse the full operative constellation or to select some transient or permanent failures. The first function executed by the tool is the orbital propagator which computes the position of all the spacecraft for the desired time. The attitude propagator (if needed) combines this information with the initial attitude configuration to determine the navigation antenna line of sight, that is the direction aligned along the signal transmitted from the payload onboard each satellite. The last function performs the coverage analysis and uses as input the position vector of each satellite and the navigation antenna line of sight, computed by the two propagators. Such function determines first the coverage area associated to the constellation, counts the number of times a point on the Earth's surface is inside the coverage region of each satellite, and according to the final sum, the four indices defined in Sec. III are computed.

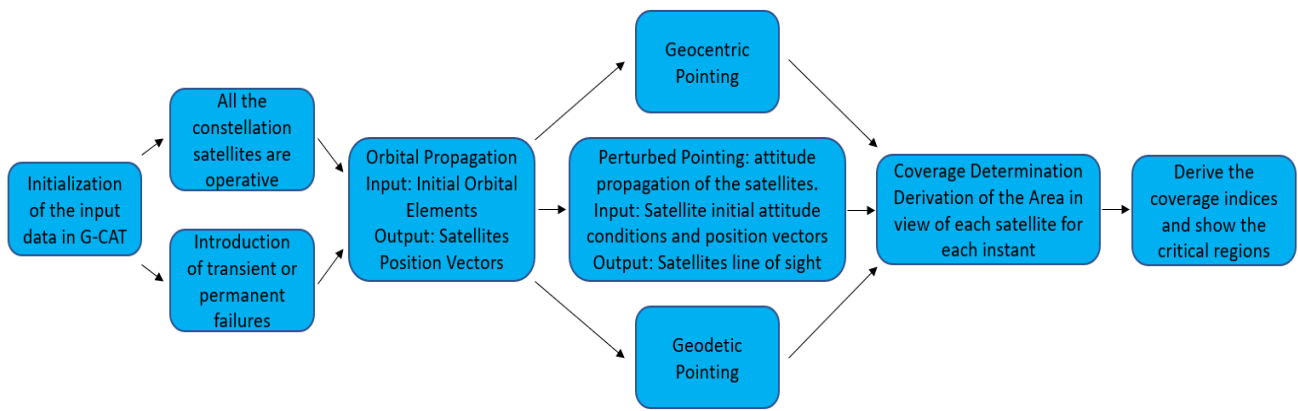

Fig. 8. Order of execution of the G-CAT functions for the GNSS coverage performance analysis

## V. GNSS Coverage Analysis and Results with the G-CAT Tool

This section presents the results obtained by the G-CAT tool when configured with the nominal operational conditions of the GPS and Galileo constellations. These results are used to validate the tool design and implementation.

In the second part of the section some failure scenarios are considered to show all the functionalities of the tool.

### V.1. GPS Nominal Configuration

The GPS is the first GNSS constellation in orbit for navigation and communication purposes. The nominal configuration is a Walker Delta pattern 30/6/1, which involves 30 satellites placed in 6 equally spaced orbital planes each one having an inclination of 55° with respect to the equatorial plane. The minimum number for obtaining GPS information is 24 [31], but the additional satellites provide global coverage with a refined precision positioning. They can also be used as spare satellite. The revolution period of each satellite is half sidereal day.

All this information is used to configure the G-CAT tool together with a geocentric pointing. The coverage performance analysis can be run for a single satellite revolution. The inputs for the coverage panel part are the following:
- Ellipsoidal representation of the Earth's surface.
- Minimum elevation angle equal to 5°.
- Geocentric pointing.

The results of the tool for the green and global coverage indices for a full operative capability of 30 and 24 satellites are shown in Fig. 9 and Fig. 10, respectively. In both the configurations, the coverage is global. However, as for the case of 24 satellites, it is not guaranteed that each point on the Earth's surface is visible by more than 4 satellites. Moreover, the additional satellites can improve the failure robustness of the GPS constellation.

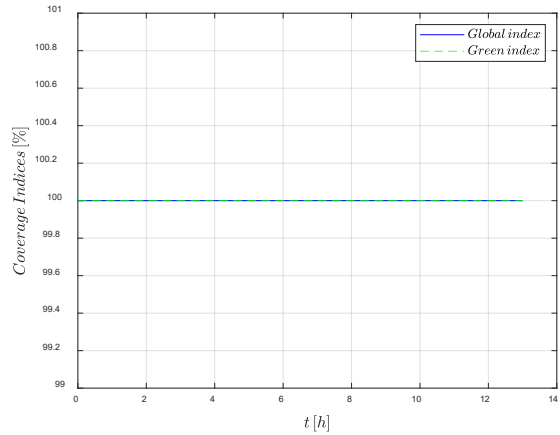

Fig. 9. Green and global coverage indices for GPS considering 30 operative satellites

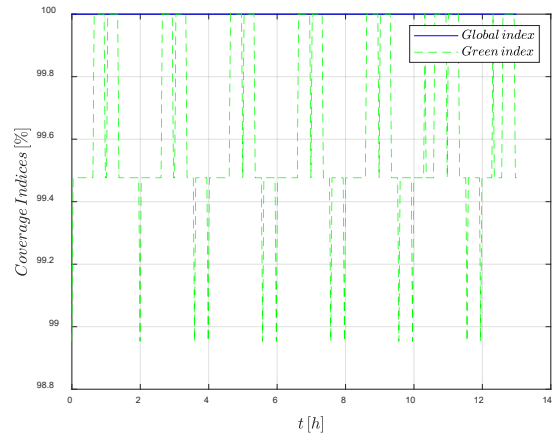

Fig. 10. Green and global coverage indices for GPS considering 24 operative satellites

### V.2. Galileo Nominal Configuration

The European GNSS Galileo is a 24/3/1 Walker Delta pattern constellation. For this reason, there are 24 satellites placed in 3 equally spaced orbital planes each one having an inclination of 56° [32]. The nominal configuration includes 6 additional satellites to be used as spare satellites in case of catastrophic failures. The revolution period of each satellite is slightly higher (15 hours) and a global coverage is required. It is possible to repeat the same procedure performed for the GPS constellation by configuring similar options for the simulation:
- Ellipsoidal representation of the Earth's surface.
- Minimum elevation angle equal to 5°.
- Geocentric pointing.

As shown in in Fig. 11, the coverage is global and there is also a higher percentage of regions on the Earth's surface visible from more than 4 satellites for an amount of time longer than the GPS case.

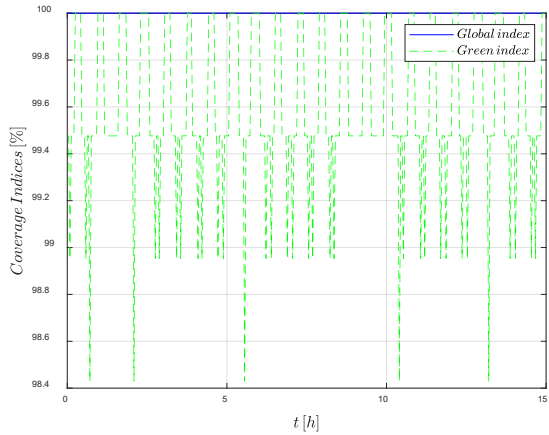

Fig. 11. Galileo coverage indices in nominal configuration

### V.3.   *Failure Case #1*

It is interesting to analyse what happens when there are off-nominal configurations. The first failure scenario involves a catastrophic failure of the first satellite of the constellation. In order to set a catastrophic failure, the option "payload safe mode" has to be flagged to simulate the transition from the nominal mode into a configuration where there is the outage of the navigation signal (see Sec. III). The Galileo case study with the following G-CAT configuration is considered:
- Ellipsoidal representation of the Earth's surface.
- Half-aperture angle equal to 12°.
- Geodetic pointing.
- Catastrophic failure of the first satellite of the constellation.

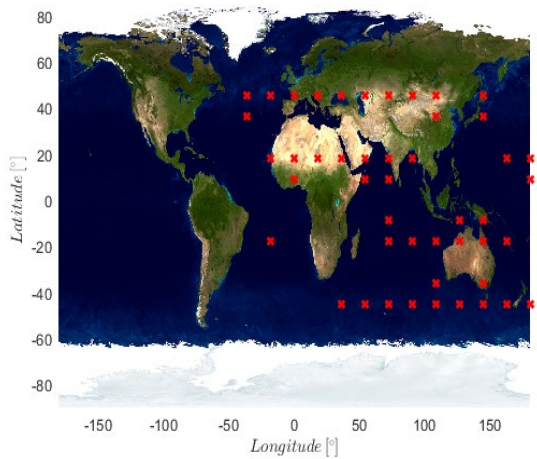

Fig. 12. 2D representation of critical regions for failure case #1

The outcome of the simulation is presented both in 2D version (see Fig. 12) and 3D version (see Fig. 14). The red crosses identify the points on the Earth's surface uncovered for at least one time instant during the whole simulation. This is the criterion for checking the global coverage for the GNSS constellation.

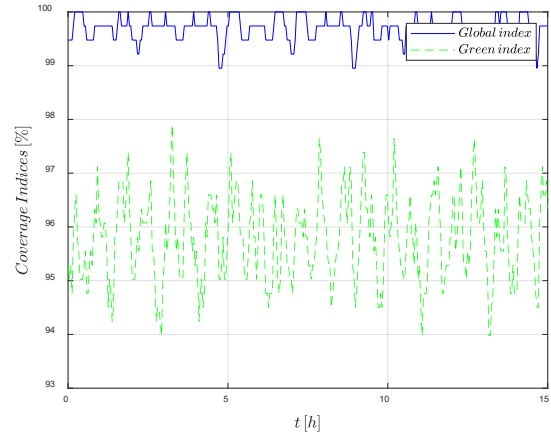

Fig. 13. Galileo coverage indices for failure case #1

Even if the coverage indices of Fig. 13 already show that a global coverage is not achieved, the 2D and the 3D representations of the terrestrial globe allow a direct association between the critical points and their geographic locations on the Earth's surface.

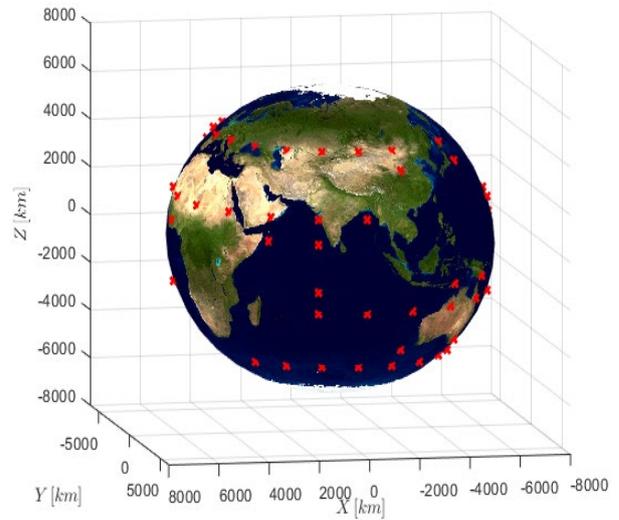

Fig. 14. 3D representation of critical regions for failure case #1

### V.4. *Failure Case #2*

The G-CAT tool allows analysing the GNSS coverage performance with both a fixed direction and a moving line of sight. In this operational scenario, the effect of the attitude perturbations in the analysis for the Galileo constellation has been included, thus the following inputs have been set in the G-CAT user interface:
- Ellipsoidal representation of the Earth's surface.

- Half-aperture angle equal to 12°.
- Perturbed pointing.
- Catastrophic failure of the first satellite of the constellation.

The results are presented in Fig. 15, where the yellow circles highlight the difference with respect to Fig. 12. Indeed, the perturbed pointing introduces differences in the coverage analysis that cannot be clearly understood by looking only at the coverage index plots because of their similarity (see Fig. 13 and Fig. 16).

Even though from an average point of view the two types of pointing seem to show similar coverage performance, it is important to use the perturbed pointing and the 2D/3D representations to identify which are the actual affected regions by the attitude perturbations.

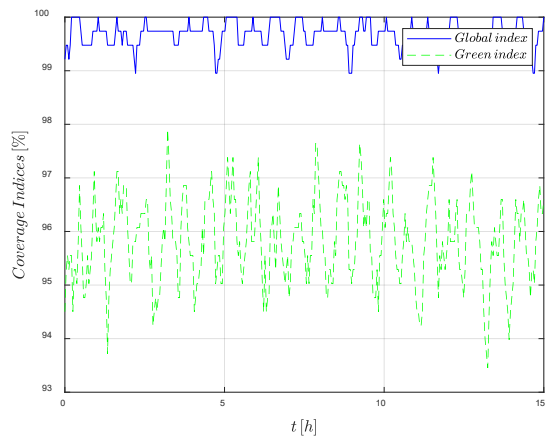

Fig. 16. Galileo coverage indices for failure case #2

The outcome of the simulation is shown in Fig. 17 and Fig. 18. As expected, the global coverage is not achieved during the failure recovery time, which is much shorter than the orbital period. This is shown by the red cross of Fig. 18.

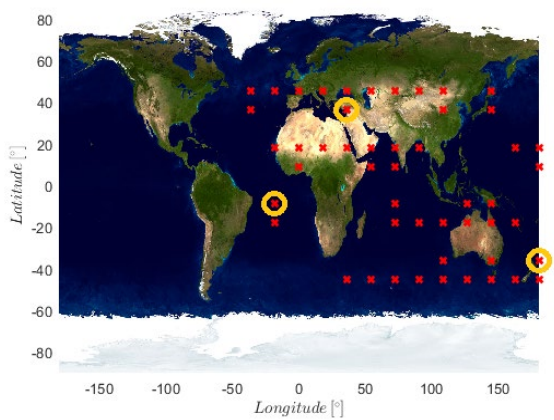

Fig. 15. 2D representation of critical regions for failure case #2. Yellow circles highlight the differences with respect to failure case #1

### V.5. Failure Case #3

The last operational case addresses the occurrence of a transient failure. For example, there is the possibility that a reconfiguration of the navigation signal makes the satellite unavailable for a limited time. The coverage performance changes for this small portion of time and even if nothing varies in terms of average percentage index, the actual coverage is affected.

In this case, in the failure panel of the tool, the "payload safe mode" has to be flagged, but this time the "recovery time" shall be less than the time used for the coverage analysis simulation. There is also the possibility to shift the time instant at which the failure begins by filling the "failure time" box.

The same set of inputs has been considered with the only difference that the first satellite of the constellation is unavailable for 20 minutes:

- Ellipsoidal representation of the Earth's surface.
- Half-aperture angle equal to 12°.
- Perturbed pointing.
- First satellite of the constellation not operative for 20 minutes.

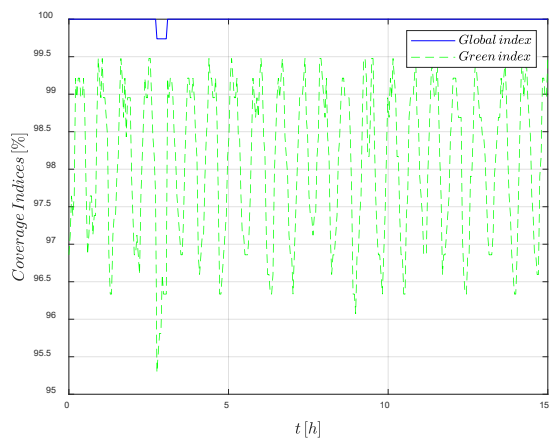

Fig. 17. Galileo coverage indices for failure case #3

The failure recovery procedure is defined by the ground stations, and it is important to execute it properly to minimise its effect on the coverage performance requirements. For instance, uncrowded regions can be addressed by such recovery procedure, instead of densely populated areas.

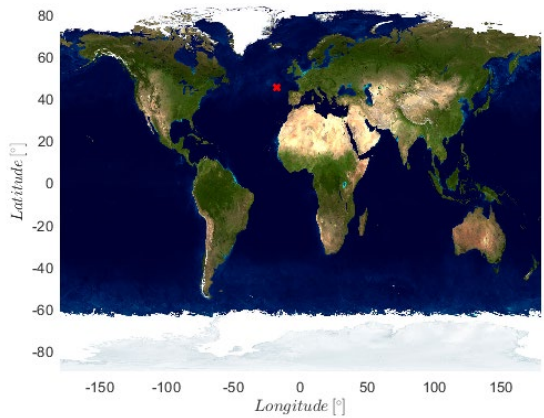

Fig. 18. 2D representation of critical regions for failure case #3

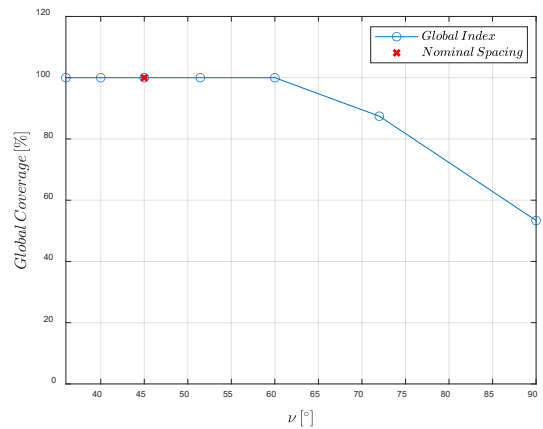

Fig. 19. Mean global coverage index for different Galileo constellation configurations

## VI. GNSS Design using G-CAT

In this section, the G-CAT tool is used to design the optimal number of satellites per orbital plane given the orbit parameters as well as the coverage and accuracy requirements. Again, Galileo has been used as reference for discussing the results provided by the tool. Basically, the design process consists in the execution of many simulations considering each time a different number of satellites. This pragmatic approach can be applied under the assumption of having the same number of satellites in all the planes with uniform angular spacing between coplanar satellites. Thanks to the low computational time associated to each simulation, it is possible to store the results of all the simulations and plot them to choose the best GNSS constellation configuration. This way, the number of satellites per orbital plane can be optimised.

The first parameter to be analysed is the global coverage index to determine the minimum number of satellites in order to achieve a global coverage. This is shown in Fig. 19, where it is clear that this minimum number is equal to 6 satellites per orbital plane associated to an angular spacing of 60°. At this stage, the obvious question to be asked is why the actual configuration of Galileo constellation involves 8 satellites per orbital plane when there are other cheaper configurations giving the same result in terms of global coverage. The answer has to be investigated in the robustness of the satellite constellation to failures. For this reason, three different failure scenarios can be considered:

- Case 1: complete loss of the first satellite of the first orbital plane.
- Case 2: complete loss of the first satellite of the first orbital plane and of the second satellite of the second orbital plane.
- Case 3: a mix of transient failures affecting the first and second satellites of the first orbital plane, the third satellite of the second orbital pane and the second satellite of the third orbital plane.

The two parameters under investigation for the optimisation process are the global index and the green index for the reasons explained hereafter.

### VI.1. Global Index Optimisation

The first parameter to be optimized is the global index. By looking at Fig. 20, it is evident that the configuration with 6 satellites per plane cannot fulfil the coverage requirements. The new optimal configuration is given by an angular spacing equal to $\nu = 51.43°$: it corresponds to 7 satellites per orbital plane. This is a proper configuration since, in all the three failure scenarios, the global index keeps close to 100%, which is the expected result to have a global coverage.

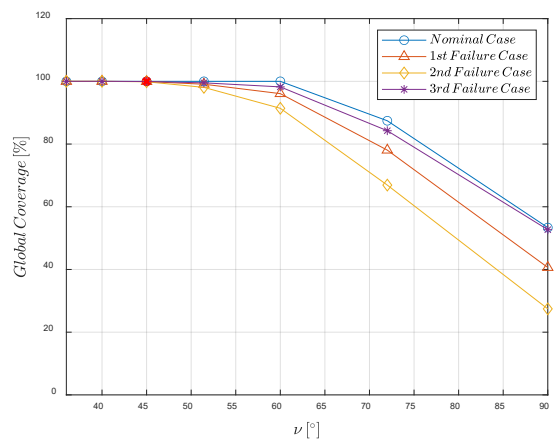

Fig. 20. Mean global coverage index for different Galileo constellation configurations

### VI.2. Green Index Optimisation

In this case, only the green index, which is the percentage of number of points on the grid representing the Earth's surface covered by more than 4 satellites, has to be optimised. The reason for this choice is related to the GNSS accuracy requirements which are usually very tight. In Sec. II it has been explained that the minimum number of satellites to calculate the user's position is equal to four. Of course, there is no possibility to improve the results in the position determination if only four measurements are provided for the trilateration process since they are associated to a unique solution. This is the reason why the green index has to be optimised in such a way that there are more satellites measurements. This gives the possibility of improving the accuracy of the determined position.

The outcome of the green index optimisation process is shown in Fig. 21, which proves that a proper configuration can be achieved with 8 satellites per orbital plane. Indeed, the previous optimal number of 7 satellites per orbital plane does not guarantee the accuracy required from the Galileo constellation.

An important remark is that Galileo original configuration involved 9 satellites per orbital plane, while the actual configuration relies on 8 operative satellites per plane. The G-CAT optimisation process outcome is in line with the actual Galileo constellation configuration, and this confirms the potentialities of using the G-CAT tool for the GNSS constellation preliminary design.

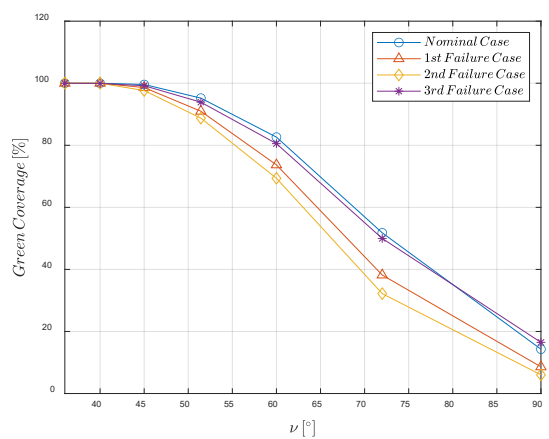

Fig. 21. Mean green coverage index for different Galileo constellation configurations

## VII. Conclusion

In this paper, a system-level engineering approach for the coverage performance analysis of a generic GNSS constellation has been presented. The corresponding tool, named GNSS Coverage Analysis Tool (G-CAT), has been designed and implemented. Such tool provides different types of pointing scenarios and gives the possibility to analyse not only the nominal configurations of the GNSS constellation, but also to verify its robustness to transient or catastrophic failures.

Each function, such as the orbit propagator, can be configured via specific panels of the G-CAT graphical user interface. Space system engineers can select the type of modelling for the Earth's surface, the pointing mode, and can add the presence of catastrophic or transient failures before performing the GNSS coverage analysis. Moreover, such tool can also be used to determine the number of satellites per orbital plane given the corresponding orbital parameters and coverage requirements. The results have been validated by considering existing GNSS constellations, such as GPS and Galileo.

Considering its current implementation status, the G-CAT tool can only be used during the initial phases of the GNSS constellation design. As for future work, the G-CAT models will be refined in order to provide space system engineers with a tool suitable for the design of future satellite constellations during more advanced project stages. In this regard, it is important to increase the total number of satellites handled by the tool, since future LEO constellations are expected to include hundreds of satellites. Another important aspect is to improve the modelling of both the ionospheric and tropospheric effects on the navigation signal propagation, such that the resulting coverage performance analysis becomes more realistic. A further refinement for both the orbit and attitude propagators is planned. The orbit propagator can provide more accurate state vectors for each satellite of the constellation, while the current ideal control law of the attitude propagator can be enhanced by modelling the real control torques provided by the spacecraft actuators (e.g., reaction wheels). More complex failure scenarios can also be implemented in order to include more difficult recovery procedures at constellation level and assess their impact on the coverage performance. Finally, it is planned to extend the G-CAT capabilities for other relevant GNSS applications, such as the ones related to the GNSS Safety-of-Life services.

## Acknowledgements

The research leading to these results has received funding from the European Research Council (ERC) under the European Union's Horizon 2020 research and innovation program as part of project COMPASS (Grant agreement No 679086). The dataset generated for this study can be found in the repository at the link www.compass.polimi.it/publications.